\documentclass[aip,graphicx,amsmath,amssymb,
 reprint]
 {revtex4-1}

\usepackage{graphicx}
\usepackage{dcolumn}
\usepackage{bm}

\usepackage[utf8]{inputenc}
\usepackage[T1]{fontenc}
\usepackage{mathptmx,amsmath}
\usepackage{etoolbox}
\usepackage{lipsum}
\usepackage{subcaption} 
\usepackage{booktabs}
\usepackage{makecell}
\usepackage{bm}

\usepackage{mylines}
\usepackage{myplots}
\newcommand{\sy}[2]{\mbox{(\kern-.25em\SymbolRGB[solid]{#1}{.8pt}{#2}{4pt}\kern-.25em)}}
\newcommand{\linesy}[3]{\mbox{(\kern-.1em\lineSymbolRGB{#1}{#2}{1.5pt}{#3}{4.5pt}\kern-.45em)}}
\newcommand{\lcap}[2]{~\,{\kern-1em\protect\mylcap{#1}{#2}}}

\definecolor{black}{RGB}{0, 0, 0}
\definecolor{blackR}{RGB}{25.5, 25.5, 25.5}
\definecolor{greyR}{RGB}{102, 102, 102}
\definecolor{redR}{RGB}{227, 47.3333, 39}
\definecolor{redRr}{RGB}{246, 191, 189}
\definecolor{greenR}{RGB}{55, 160.3333, 85}
\definecolor{greenRr}{RGB}{194, 225, 203}
\definecolor{blueR}{RGB}{55, 135, 192.3333}
\definecolor{blueRr}{RGB}{194, 218, 235}
\usepackage[normalem]{ulem}
\usepackage{color}

\definecolor{bluecomment}{RGB}{20, 20, 255}

\makeatletter
\def\@email#1#2{%
 \endgroup
 \patchcmd{\titleblock@produce}
  {\frontmatter@RRAPformat}
  {\frontmatter@RRAPformat{\produce@RRAP{*#1\href{mailto:#2}{#2}}}\frontmatter@RRAPformat}
  {}{}
}%
\makeatother

\begin{document}


\title[Machine learning flow control with few sensor feedback and measurement noise]{Machine learning flow control with few sensor feedback and measurement noise}
\author{R. Castellanos$^*$}  \email{rcastell@ing.uc3m.es}
\affiliation{Aerospace Engineering Research Group, Universidad Carlos III de Madrid, Legan\'es 28912, Spain}
\affiliation{Theoretical and Computational Aerodynamics Branch, Flight Physics Department, Spanish National Institute for Aerospace Technology (INTA), Torrej\'on de Ardoz 28850, Spain}

\author{G. Y. Cornejo Maceda}
\affiliation{School of Mechanical Engineering and Automation, Harbin Institute of Technology (Shenzhen),  University Town, Xili, Shenzhen 518055, People’s Republic of China}

\author{I. de la Fuente}
\affiliation{Aerospace Engineering Research Group, Universidad Carlos III de Madrid, Legan\'es 28912, Spain}%

\author{B. R. Noack}
\affiliation{School of Mechanical Engineering and Automation, Harbin Institute of Technology (Shenzhen),  University Town, Xili, Shenzhen 518055, People’s Republic of China}
\affiliation{Institut für Strömungsmechanik und Technische Akustik (ISTA), Technische Universität Berlin, Müller-Breslau-Straße 8, D-10623 Berlin, Germany}

\author{A. Ianiro}
\affiliation{Aerospace Engineering Research Group, Universidad Carlos III de Madrid, Legan\'es 28912, Spain}%

\author{S.Discetti}
\affiliation{Aerospace Engineering Research Group, Universidad Carlos III de Madrid, Legan\'es 28912, Spain}%

\date{\today}

\begin{abstract}
A comparative assessment of machine learning (ML) methods for active flow control is performed. The chosen benchmark problem is the drag reduction of a two-dimensional Kármán vortex street past a circular cylinder at a low Reynolds number ($Re=100$). 
The flow is manipulated with two blowing/suction actuators on the upper and lower side of a cylinder. The feedback employs several velocity sensors. Two probe configurations are evaluated: 5 and 11 velocity probes located at different points around the cylinder and in the wake. 
The control laws are optimized with Deep Reinforcement Learning (DRL) and Linear Genetic Programming Control (LGPC). 
By interacting with the unsteady wake, both methods successfully stabilize the vortex alley and effectively reduce drag while using small mass flow rates for the actuation. DRL has shown higher robustness with respect to variable initial conditions and to noise contamination of the sensor data; on the other hand, LGPC is able to identify compact and interpretable control laws, which only use a subset of sensors, thus allowing reducing the system complexity with reasonably good results. 
Our study points at directions of future machine learning control combining desirable features of different approaches.

\vspace{0.2cm}
\end{abstract}

\maketitle 

\section{Introduction}\label{s:Introduction}

Flow control of turbulent flows, in particular with the purpose of drag reduction, is a recurrent research objective that has regained interest during the last decade \citep{Noack2019control}. Machine learning, in particular, plays a key role in the development of sophisticated and efficient flow-control algorithms \citep{BruntonNoackKoumoutsakos2020} and presents a possible solution to the difficulties imposed by the non-linearity, time-dependence, and high dimensionality inherent to the Navier-Stokes equations.

Control of wake flows, in particular, has attracted significant interest due to its relevance in a wide variety of research and industrial applications. Control strategies commonly target two main types of drag: the skin friction drag, caused due to the viscosity of the fluid interacting with the wall surface, and the wake drag, which originates after the affected body.

Passive drag reduction methods, such as the classical dimples on the surface of a sphere \citep{Bearman1993dimples} to delay transition or the use of splitter plates in the wake \citep{Ozono1999splitter} to suppress the vortex shedding, have shown to be quite successful. Nonetheless, during these last decades, hardware/software advances are pushing towards active methods instead, which exploit the potential advantage of tuning the action according to the flow state. 

Active drag reduction techniques initially focused on increasing base pressure to reduce pressure drag, using transpiration and vibration techniques for such an objective. Continuous or pulsating-based-bleeds could be used to modify the flow in the separated region. For the latter, drag reduction could be achieved with zero net mass addition, with maximum effectiveness at a frequency twice the K\'arm\'an shedding frequency \citep{Williams1989}. Among the wide variety of active drag reduction techniques that resulted effectively, small jets have shown to be very efficient, enabling separation control with weak actuation \citep{glezer2011control}. 

The design of effective flow control techniques is a challenging objective, especially when the solution is based only on limited velocity or pressure data extracted from the fluid flow \citep{duriez2017book}. In its essence, the flow control problem can be described as a functional optimization problem, in which the state of the dynamical system has to be inferred from a limited number of observable quantities. The objective is to find a control function that minimizes (or maximizes) a cost (or reward) function, based on the desired features for the controlled flow configuration.

Considering the categorization of control strategies, one of the main classifications is based on the existence of a model to describe the system to be controlled, distinguishing between model-based and model-free control. The latter encloses the kind of control strategies where an optimized control law is extracted without imposing any model of the dynamical system. The popularity of these approaches has been growing considerably in the last decade, thanks to the popularization and advances in machine learning techniques. Two of the most prominent model-free control techniques from the machine learning literature are Reinforcement Learning \citep[RL]{sutton2018reinforcement} and Genetic Programming \citep[GP]{koza1994}. Reinforcement Learning is an unsupervised learning technique focused on optimizing a decision-making process interactively, which makes it a preferred choice in flow control over its alternatives. On the other hand, Genetic Programming algorithms are focused on recombining good control policies by systematic testing, exploiting the ones with the best results and exploring possible alternatives in the solution space. 

Genetic Programming, originally pioneered by \citet{koza1994} belongs to the family of Evolutionary Algorithms (EA), which have a common workaround: a population of individuals, called a generation, compete at a given task with a well-defined cost function, and evolve based on a set of rules, promoting the most successful strategies to the next generation \citep{banzhaf1998genetic, duriez2017book}. It constitutes a powerful regression technique able to re-discover and combine flow control strategies, which have been proven useful in the cases of multi-frequency forcing, direct feedback control and controls based on ARMAX (Autoregressive Moving Average eXogenous), without any physics information \citep{cornejomaceda2019pamm,cornejo2021gMLC}. 
Machine Learning Control (MLC)\citep{duriez2017book} based on tree-based Genetic Programming has been able to develop laws from small to moderate complexity, e.g. the phasor control, threshold-level based control, periodic or multi-frequency forcing, including jet mixing optimization with multiple independently unsteady minijets placed upstream of nozzle exit \citep{zhou_artificial_2020}, the analysis of the effect of a single unsteady minijet for control \citep{wu_jet_2018}, drag reduction past bluff bodies \citep{li2019prf} , shear flow separation control \citep{gautier2015MLC}, reduction of vortex-induced vibrations of a cylinder \citep{Ren2019pof}, mixing layer control \citep{Parezanovic2016jfm}, and wake stabilization of complex shedding flows \citep{raibaudo2019pof} among others.
Some relevant improvements have been made to the MLC framework  such as the integration of a Linear-based Genetic Programming (LGP) algorithm \citep{li2017GP}, which is the chosen option in this study. Recently, a faster version of MLC based on the addition of intermediate gradient-descent steps between the generations (gMLC) has been developed \citep{cornejo2021gMLC}

Reinforcement Learning is based on an agent learning an optimized policy based on the different inputs and outputs. The key feature of RL is that the only information available for the algorithm is given in the form of a penalty/reward concerning a certain action performed by the model, but no prior information is available on the best action to take. In Deep RL (DRL), the agent is modelled by an Artificial Neural Network (ANN) which needs to be trained \citep{rabault2019DRL}. ANN  have a long history of use to parametrize control laws \citep{Lee1997NN}, or to find the optimized flow control strategy for several problems such as swimming of fish schoolings \citep{gazzola2014RL}, control of unmanned aerial vehicles \citep{bohn2019DRL}, and optimization of glider trajectory taking ascendance \citep{reddy_learning_2016}. On the other hand, the exploitation of DRL specifically on flow control is relatively recent. The first applications targeted the control of the shedding wake of a cylinder in simulations \citep{rabault2019DRL,rabault2019JFM} and in experiments \citep{Fan2020}. 
DRL has been enforced in tuning the heat transport in a two-dimensional Rayleigh–Benard convection \citep{Beintema2020}, in the control of the the interface of unsteady liquid films \citep{Belus2019} and in stabilizing the wake past a cylinder by imposing a rotation on two control cylinders located at both sides \citep{Xu2020joh}. 
Recently, \citet{li2021ReLe} investigated how to use and embed physical information of the flow in the DRL control of the wake of a confined cylinder, and \citet{paris2021} explored the utilization of DRL for optimal sensor layout to control the flow past a cylinder.

It can be argued that LGP and DRL share many similarities, up to the point in which many fitness functions in LGP can be considered as DRL systems\citep{Banzhaf1997}. The recent ongoing developments of DRL and LGP in flow control applications open up relevant questions on their applicability in experimental environments, where the number of sensors is limited and data are likely to be corrupted by noise. Furthermore, the generalization of the identified policies is often hindered by the challenges of interpreting the control laws. This work sheds light on the main features of Deep Reinforcement Learning as Linear Genetic Programming Control (LGPC) in this direction. To the author's best knowledge, only the recent contribution by \citet{pino2022comparative} performs a comparative assessment of machine learning methods for flow control. The work focuses on the comparison of the relation of DRL and LGPC with optimal control for a reduced set of sensors. The robustness to noise and the effect of initial condition of such algorithms is not discussed. The present study aims to address the performance of DRL and LGPC in the simple scenario of the control of the 2D shedding wake of a cylinder at a low Reynolds number in the conditions of a limited number of sensors. The robustness of both processes to noise contamination on the sensor data and variable initial conditions for training individuals is assessed. Finally, an interpretation of the control actions using a cluster-based technique is provided. For this purpose, the same DRL framework and simulation environment used by \citet{rabault2019DRL} is considered, and compared against the LGPC environment developed by \citet{li2017GP}.
It is to be noted that this contribution is a proof of concept. The simulation environment proposed by \citet{rabault2019DRL} was chosen given its simplicity and affordability for extensive analysis as herein presented. Nonetheless, the application of akin algorithms to similar environments, though with more challenging conditions, was investigated by \citet{Tang2020pof}, achieving a robust control in the flow past the confined cylinder at multiple Reynolds numbers, concluding that the drag reduction increases with the Reynolds number. Additionally, \citet{Ren2021pof} exploits the framework by \citet{rabault2019DRL} at weak turbulent conditions (Re = 1000) with a drag reduction of 30\%. Nonetheless, in these studies, 236 and 151 probes are considered respectively, and the robustness to noise is not explored.

The present article is structured as follows: Section \ref{s:Methodology} defines the main methodologies applied to implement the different machine learning models, the simulation environment and the problem description. Results are collected and described in Section \ref{s:result} while the interpretation of the achieved controls and their performance is outlined in Section~\ref{s:interpretability}. Ultimately, the conclusions of the study are drawn in Section \ref{s:Conclusions}.

\begin{figure*}[t]
    \centering
    \includegraphics[width=0.9\linewidth]{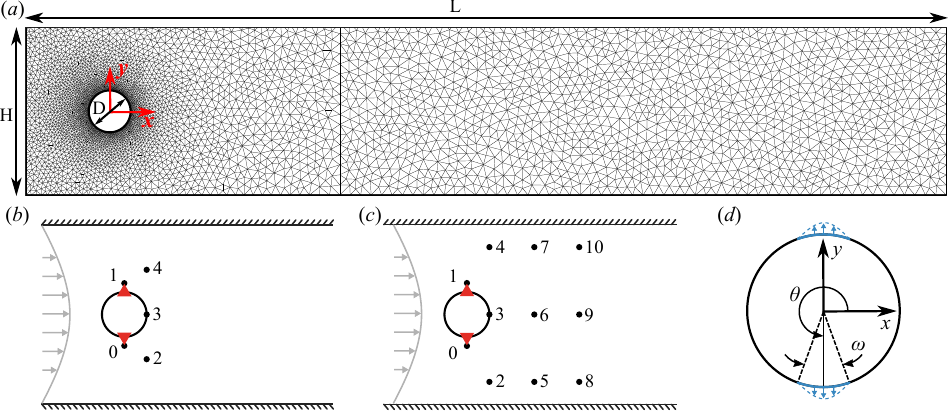}
    \caption{Description of the numerical setup adapted from \citet{rabault2019JFM}. (a) Sketch of the numerical domain and the non-structured mesh, defining the diameter $D=1$, length $L=20$, and height $H=4.1$. The sensor arrangement is shown for (b) 5 and (c) 11 probes, being probes identified as \sy{black}{o*} and actuators as \sy{redR}{t*}. (d) Detail of the cylinder and the jet actuators definition.}
    \label{fig:NumericalSetup}
\end{figure*}

\section{Methodology}\label{s:Methodology}

This Section is focused on the main methodologies to define the environments in which the Machine Learning models will be tested, as well as the conditions imposed. Although the methodologies to be compared present important differences, they share several fundamental characteristics that must be highlighted before delving into each of the independent methods.

\subsection{Simulation Environment}

The active drag reduction is performed in a 2D Direct Numerical Simulation (DNS) environment that builds upon that of \citet{rabault2019JFM}, differing only for the sensor strategy. The environment is described here for completeness. 

The geometry of the simulation, adapted from state of the art benchmarks \cite{schafer_benchmark_1996}, consists of a cylinder of diameter $D$ immersed in a box of total length $L=22D$ (along the $x$-axis) and height $H=4.1D$ (along the $y$-axis) as shown in figure~\ref{fig:NumericalSetup}(a). The inflow velocity (on the left wall of the domain) is modelled as a parabolic profile so that the mean velocity magnitude results in $U$.
A no-slip boundary condition is imposed on the top and bottom walls of the channel, and also on the solid cylinder walls. An outflow Dirichlet boundary condition is imposed on the right wall of the domain. The Reynolds number based on the mean velocity magnitude and cylinder diameter ($Re=\frac{{U} D}{\nu}$, with $\nu$ the kinematic viscosity) is set to $Re=100$. This choice is based on the previous work by \citet{rabault2019JFM} which has been a pioneer application
of DRL to flow control and the baseline to compare within this study. Working with higher Reynolds numbers would have implied a tremendous increase in computational cost, making unaffordable all the set of simulations required for changing the initial conditions and assessing the noise robustness that will be discussed in the following.

The control action is performed by two jets (1 and 2) controlled through a non-dimensional mass flow rate $Q_{\rm jet}$ by imposing a parabolic velocity profile with the jet width of $\omega = 10^\circ$. The jets are perpendicular to the cylinder wall and located at angles $\theta_1 = 90^\circ$ and $\theta_2 = 270^\circ$ relative to the flow direction as shown in figure~\ref{fig:NumericalSetup}(d), what guarantees that all the promoted drag reduction is the result of indirect flow control, rather than direct injection of momentum~\citep{rabault2019JFM} To prevent numerical instability while presenting a more realistic scenario, the total mass flow rate injected by the jets is zero, i.e. $Q_{\rm jet_1} = -Q_{\rm jet_2}$. Note, however, that the cylinder does not present physical cavities on its surface, meaning that there is no physical interference of the jet slot with the flow field.

The simulation environment is based on the open-source finite-element framework FEniCS \cite{logg2012book} version 2017.2.0, solving the unsteady Navier–Stokes equations equations by DNS.  Computations are performed on an unstructured mesh generated with Gmsh \cite{geuzaine2009gmsh}. The mesh is refined around the cylinder and is composed of $9262$ triangular elements (see figure~\ref{fig:NumericalSetup}(a)). A non-dimensional, constant numerical time step $dt=5 \times 10^{-3}$ is used. The CFL condition is enforced in the problematic zones, that is, close to the actuation jets, by imposing a maximum jet mass flow rate ($|Q_{\rm jet}|<Q_{\rm jet_{max}}$). 

The flow control framework developed by \citet{rabault2019JFM} was conceived to work either with velocity or pressure probes as sensing. Since pressure probes are often more difficult to be installed in a customary location for an experimental application, it was preferred to chose the velocity probes which resembles what could be extracted from hot wire anemometry (HWA). Two sensor configurations are considered in the present study with $5$ and $11$ probes which report the local value of the horizontal and vertical components of the velocity field (see figure~\ref{fig:NumericalSetup}(b) and \ref{fig:NumericalSetup}(c), respectively). The probes are located in the wake of the cylinder to enable the controller to learn from the vortex shedding pattern.

\subsection{Formulation of the optimization problem}

\begin{figure*}
    \centering
    \includegraphics[width=0.9\linewidth]{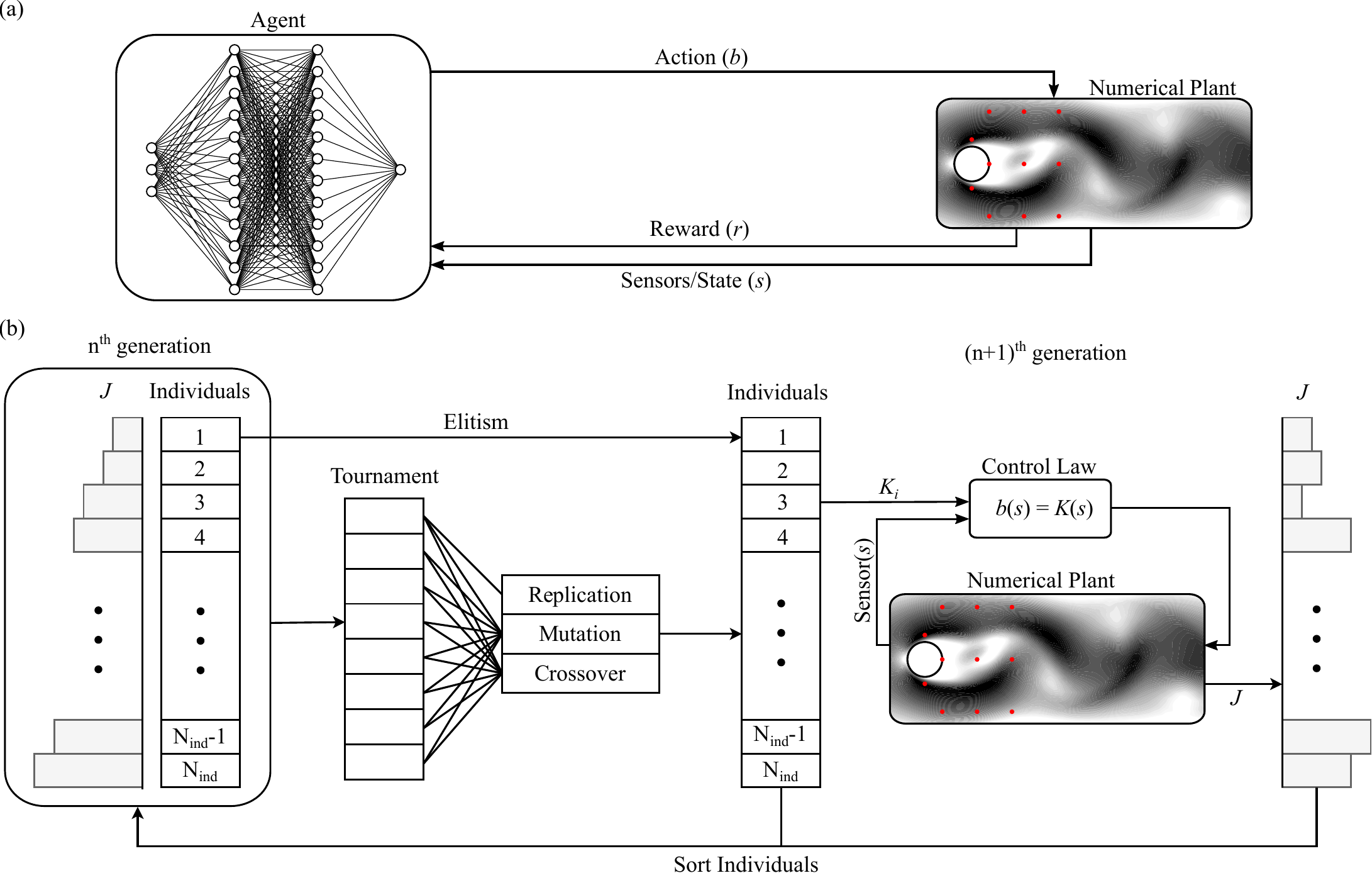}
    \caption{Implementation of the control algorithms. (a) Sketch of the closed-loop control based on Deep Reinforcement Learning. (b) Sketch of the closed-loop control based on Linear genetic Programming.}
    \label{fig:LearningLoop}
\end{figure*}

The drag reduction of a 2D cylinder wake flow is a classical optimization problem with a simple target, i.e. reducing drag. The control problem is formulated as a regression problem, i.e. to find the control law which optimizes a given cost function $J$ \citep{duriez2017book}. The proposed cost function has been shaped as a combination of drag and lift coefficients, modifying the one proposed by \citet{rabault2019JFM} as follows:

\begin{equation}
    J= 1+\left\langle C_{D}\right\rangle_{T} - \left\langle C_{D_0}\right\rangle_{T}+0.2\left|\left\langle C_{L}\right\rangle_{T}\right| \label{Eq:J}
\end{equation}

where $\langle C_{D_0}\rangle_T = 3.206$ is the drag coefficient of the unforced flow, and $\langle \cdot \rangle_T$ indicates the sliding average back in time over a duration corresponding to one vortex shedding cycle $T$ of the unforced vortex shedding flow. 

The cost function in equation~\ref{Eq:J} has shown to be better than using just the instantaneous drag coefficient, i.e. $J(t) = C_D(t)$. First, the average values of the lift and drag coefficients over one vortex shedding cycle reduce the oscillations of the objective function, which has been reported to improve learning speed and stability \citep{rabault2019JFM}. Secondly, the drag reduction is defined as the increment with respect to the unforced flow ($\Delta C_D = \left\langle C_{D}\right\rangle_{T} - \left\langle C_{D_0}\right\rangle_{T}$), which is intended to be as negative as possible. Thirdly, a penalization term based on the lift coefficient is considered to prevent the controller from finding undesired asymmetric solutions\citep{rabault2019JFM}. Finally, adding the unity to the cost function provides a bias to prevent $J$ negative values that could affect the convergence of the learning process. Ultimately, the algorithm would try to minimise $J$, by reducing drag while maintaining low lift components.

From an optimization perspective, the resultant control is the strategy that minimizes the cost function with a control law $\bm{b}(t) = {K}(\bm{s}(t))$, where $\bm{b}(t) = (b_1(t), ...,b_{N_b}(t))^T$ comprises $N_b$ actuation commands, and $\bm{s}(t) = (s_1(t), ..., s_{N_s}(t))^T$ consist of $N_s$ sensor signals. In this study, the actuation $b$ is performed with two jets that are related due to the net zero flux condition ($N_b = 1$) at the bottom and top sides of the cylinder, using velocity probes ($N_s = 5$ or $11$) as sensing. The control problem is equivalent to finding ${K}^{*}$ such that
\begin{equation}\label{eq:Optproblem}
    \begin{aligned}
        {K}^{*}(\bm{s}) = \underset{{K}}{\arg\min}~~ J({K}(\bm{s}))
    \end{aligned}
\end{equation}

The optimized feedback ${K}^{*}$ is computed following LGPC and DRL, as described in the following sections. The resultant control laws map $N_s$ sensor signals into $N_b$ actuation commands.
The resultant control action is subjected to a smoothing operation to ensure continuous control signals without abrupt alterations in the pressure and velocity due to the use of an incompressible solver. The control action is then adjusted from one-time step in the simulation to the next by
\begin{equation}\label{eq:Qsmooth}
    Q_{\rm jet}(t)=Q_{\rm jet}(t-dt)+\alpha\left[{b}(t)-Q_{\rm jet}(t-dt)\right],
\end{equation}
being $\alpha=0.1$ a numerical parameter, $Q_{\rm jet}(t)$ the jet actuation used by the plant at time instant $t$, and ${b}(t)$ the actuation proposed by the machine learning model at time instant $t$. 

\subsection{Deep Reinforcement Learning}

Deep Reinforcement Learning (DRL) is an ML control algorithm in which an agent (managed by an ANN) learns the best control by interacting with the environment to exchange information in a closed-loop process (see figure~\ref{fig:ApproxMethod}(b)). In the present work, the plant or environment is the above-mentioned simulation environment, which interacts with the agent by three channels: the observation or sensor state, $\bm{s}$ ($N_s$ point measurements of velocity); the action, $b$ ($N_b$ values of mass flow rate to impose on the jets); and the reward, $r$ (the cost function in equation \ref{Eq:J} based on $C_D$ and $C_L$, i.e. $r=J$). Based on the sensing data and the reward of the current state, DRL trains an ANN to find the optimized closed-loop control strategies that maximize the expected reward.

The DRL framework is the same as in \citet{rabault2019JFM} in which the agent uses the proximal policy optimization (PPO) method \citep{schulman_proximal_2017} for performing learning. The PPO method is episode-based, so that it iteratively learns by applying a certain control for a limited amount of time (episode duration) before analysing the rewards and sensors, and resuming learning with a new episode. The considered architecture for the ANN is relatively simple, being composed of two dense layers of $512$ fully-connected neurons, the input layer to acquire data from the probes, and the output layer to generate data for the two jets. For more details, readers are referred to \citet{rabault2019JFM}. The training loop of DRL is sketched in Figure \ref{fig:LearningLoop}(a).

At the beginning of the learning process, the PPO explores purely random controls to assess the values of the reward function. This initial approach implies difficulties to learn the necessity to set time-correlated, continuous control signals. To solve this issue, \citet{rabault2019JFM} implemented the agent such that the control value provided by the network is kept constant for a duration of $50$ numerical time steps. Therefore, the PPO agent interacts and updates the ANN coefficients only every $50$ time steps, which is the duration of a fixed actuation. This numerical trick together with the smoothing described in equation \ref{eq:Qsmooth} provides a continuous control signal.

\subsection{Linear Genetic Programming}
Linear Genetic Programming \citep[]{Wahde2008book} is an evolutionary algorithm, that applies biological-inspired operations to select the fittest individuals for a given purpose. The control laws are effectively mappings between the outputs (sensor information) and the inputs (actuation) of a dynamical system. In the following, the control laws are also referred to as \textit{individuals} to comply with the evolutionary terminology. LGP is able to learn control laws in a model-free meaner, optimizing both the structure of the function and its parameters. In practice, the control laws are internally represented by a matrix encoding a list of instructions. Each row of the matrix codes for a mathematical operation from a set of input registers, constants and operations and stores the result in a memory register. The matrix is then read linearly modifying sequentially the memory registers, hence the name of the method. The control law is then read in the first register. LGPC is here selected as the preferred option for its simpler implementation of the genetic operators for Multiple-Input-Multiple-Output control \citep{cornejomacedaPhD}.

The learning process, sketched in Figure \ref{fig:LearningLoop}(b), is divided into an outer loop devoted to evolving the generations and an inner loop to evaluate all the individuals in a real-time control process. First, an initial population of individuals is randomly generated and evaluated with a Monte-Carlo optimization to explore the control law space. We recall that the individuals are analytical functions of the input data, i.e. the velocity sensor signals. A measure of the performance of each individual is given by its cost $J$ (equation~\ref{Eq:J}). Once the entire population is evaluated, the next generation of individuals is created with genetic operators (crossover, mutation, replication) applied to the most performing individuals. The best individuals are selected with the tournament selection method. Crossover combines two individuals and generates a new pair of individuals by exchanging randomly their instructions. This operation contributes to the exploitation of the learned data by recombining well-performing individuals. The mutation operation modifies randomly elements of one given control law to explore potentially new and better minima. Replication is the memory operator. It assures that good structures are not lost in the evolution process. Finally, an elitism operation saves the best individuals of one generation to the next, ensuring that the performance does not degrade after each generation. The genetic operators (crossover, mutation and replication) are chosen following respective probabilities ($P_c,P_m,P_r$). The process is repeated for every new generation until the stopping criterion is met or if the termination is triggered. In this study, all the training processes have been performed for a fixed number of generations $N_{g} = 15$ as explained later.  

Among the wide variety of custom settings when dealing with LGPC, the most relevant parameters for proper performance and convergence of the algorithms are the population size, the number of generations, the tournament selection size and the genetic operators' probability ($P_c,P_m,P_r$). LGPC parameters are chosen following the recommendations of \citet{duriez2017book} and \citet{li2017GP}. They are summarized in table~\ref{tab:GAparameters}

\begin{table}[h]
    \centering
    \begin{tabular}{lc}
    \toprule
    Number of controllers                       & 1 \\
    Number of sensors                           & 5,11 \\
    Population size                             & 100 \\
    Number of generations                       & 15 \\ 
    Tournament selection size                   & 7   \\ 
    Crossover probability                       & 0.6 \\ 
    Mutation probability                        & 0.3 \\
    Replication probability                     & 0.1 \\
    Elitism                                     & 1   \\
    Operations                                  & $+$, $-$, $\times$, $\div$, $\sin$, $\cos$, $\tanh$ \\ \bottomrule
    \end{tabular}
    \caption{Selection of LGPC parameters} \label{tab:GAparameters}
\end{table}

An important difference between the DRL and the LGPC algorithm is related to their learning process during an episode. In DRL, the agent builds its internal representation of how the flow in a given state will be affected by actuation, and how this will affect the reward value. This is done globally at the end of the episode and also each time the actuation changes. The agent is modified according to the expected reward (total or partial, respectively), which is not an immediate value just after the actuation, but also after the medium/long-term reward. This means that the neural network on which the DRL is capable of learning both from the committed errors but also from the future effect associated with each actuation. In LGPC, there is no chance of the actuation law during the simulation run, as it corresponds to a single individual mapping from the input to the outputs. It is therefore interesting considering the incorporation of time-delayed sensor signals. Following \citet{cornejo2021gMLC}, the values of the probes at $1/4, 1/2$ and $3/4$ of the shedding period are included, as well as the instantaneous value. Assuming a periodic flow, the addition of such time delays enables the reconstruction of the flow phase.

\subsection{Training standards}
The training process of both DRL and LGPC is performed with the same parameters, which were chosen based on the recommendations by \citet{rabault2019JFM} and extensive empirical analysis. The duration of the simulation (or episode duration, according to DRL nomenclature) is set to $T_{sim} = 20.0$, which translates into approximately 6 vortex shedding periods, and corresponds to 4000 numerical time steps. Note, however, that the cost function in equation~\ref{Eq:J} is evaluated for the last shedding period (650 numerical time steps) in which a new steady state is expected to be reached upon actuation.

Regarding the action, the DRL agent adjusts the policy every $50$ numerical time steps, which means that the control is updated $80$ times during the episode. The transition from the current action to the updated following action is smooth and continuous based on the smoothing described in equation~\ref{eq:Qsmooth}. On the other hand, LGPC provides analytical control laws that are continuous and fully dependent on the sensing data, which means that the control action is adjusted every numerical time step mildly thanks to the smoothing.

Given the intrinsic differences between LGPC and DRL, it is required to set the training standard to guarantee a fair comparison. The followed criterion is to keep the same computational time during the training process. The PPO agent is able to learn a fully-stabilized control after approximately $400$ epochs (corresponding to $32000$ sample actions); however, the convergence rate is lower when applying noise to the probes. On the other hand, for LGPC with a pool size of $100$ individuals per generation, a converged control law is achieved before the $10^{th}$ generation both with and without noise consideration. Both for LGPC and DRL, it is straightforward to assess that the main computational cost comes from the plant, i.e. from the fluid mechanic simulation of the 2D cylinder wake. The evaluation of the sensing, update of the agent or generation of new individuals are operations with negligible time consumption. Based on these figures of merit, it was decided to set the training duration to $1500$ episodes in the case of DRL and $15$ generations of $100$ individuals in the case of LGPC. This common criterion guarantees that the computational effort is the same for both algorithms since a total of $1500$ simulations are launched for each method.

For the investigation of robustness to noise, a perturbation with Gaussian distribution is added to the probes such that the input used by the DRL agent or the LGPC control laws are altered. The noise implementation is the following,
\begin{equation}
    u_i(t) = u_i(t) + \varepsilon \cdot \digamma_i(t) \qquad \forall \quad i = [1,2, ... , N_b],
\end{equation}

being $u_i(t)$ the velocity value, $\digamma_i(t)$ a random normally distributed value for the probe $i$ at time instant $t$, and $\varepsilon$ the noise level or intensity. Three noise levels are considered, i.e. $1\%,5\%$ and $10\%$ of the freestream velocity. On the contrary, the mean state quantities used to compute the cost function (i.e., $C_D$ and $C_L$) are not altered by noise since the averaging operator in the cost function would make the noise of minor relevance. 

\subsection{Control law visualization}\label{Sec:Method_ControlLawVisu}
\begin{figure}
    \centering
    \includegraphics[width=\linewidth]{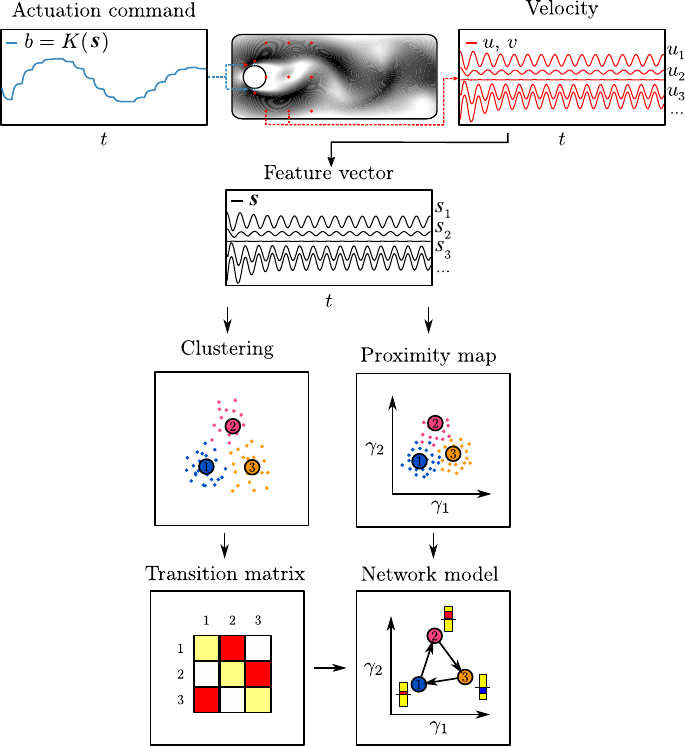}
    \caption{Control interpretation methodology. The sensor vector is built from the vertical and horizontal components of the measured velocity. The elements of the sensor vector are grouped in clusters (three clusters (1,2,3) are presented here for clarity). High probability transitions are depicted with darker colours. $\gamma_1$ and $\gamma_2$ define the projection plane for the proximity map. The actuation magnitude is represented by rectangles in the network model; red (blue) for a positive (negative) actuation with respect to the mean value. See the text for more details.}
    \label{fig:ApproxMethod}
\end{figure}

A cluster-based interpretation method is considered to have an insight into the actuation mechanisms involved in the control learned by DRL and LGPC. Cluster-based methods have been recently applied to build network models able to reproduce the main characteristics of fluid flows and dynamical systems, such as temporal evolution and fluctuation levels \citep{Fernex2021,LiH2021jfm}.

Understanding the relationship between the control inputs and the corresponding actuation command is not an easy task, especially with a large number of inputs. Thus, clustering is employed to extract representative states from the sensors. Figure~\ref{fig:ApproxMethod} summarizes the main steps of the control interpretation methodology described below. For this study, the metric employed for the cluster analysis is the one induced by the $L^2$ norm. The sensor time series are reduced to $10$ clusters. For each cluster, its centroid is computed as the average of all states in the cluster. Then, the $10$ centroids represent the main states of the flow. Moreover, the transition information from one flow state to another is gathered in a probability transition matrix ($P=[p_{i,j}]_{i,j}$) that translates the probability to jump from one cluster to another. The probability transition from cluster $i$ to cluster $j$ is defined by $p_{i,j}={n_{i,j}}/{n_i}$, being $n_{i,j}$ the number of states from cluster $i$ that transition to cluster $j$ and $n_i$ the total number of states in cluster $i$.

The centroids combined with the transition matrix allow for building a network model of the controlled flow based only on the controller inputs. On the other hand, the sensor time series are projected on a 2D proximity map with classical multidimensional scaling \citep[MDS]{Kaiser2017ifac,LiA2022jfm,foroozan2021unsupervised}. MDS is a powerful tool for dimensionality reduction, that projects the data in the two directions ($\gamma_1$,$\gamma_2$) of maximum dispersion of the feature vector distance matrix. The distance matrix is computed with the same metric as the clustering. Combining the network model and the proximity map allows to have a reconstruction of the flow phase space. Finally, the average actuation performed in the cluster is associated with each centroid. The resulting 2D visualization represents the dynamics of the flow alongside the actuation performed allowing an easy interpretation of the control actuation mechanism.


\begin{table*}[t]
\centering
\begin{tabular}{>{\centering}p{0.125\linewidth-2\tabcolsep}
                >{\centering}p{0.125\linewidth-2\tabcolsep}
                >{\centering}p{0.125\linewidth-2\tabcolsep}
                >{\centering}p{0.125\linewidth-2\tabcolsep}
                >{\centering}p{0.125\linewidth-2\tabcolsep}
                >{\centering}p{0.125\linewidth-2\tabcolsep}
                >{\centering\arraybackslash}p{0.125\linewidth-2\tabcolsep}}
\toprule
Algoritm & Probes & Noise & $\langle J \rangle$ & $\langle C_D \rangle$ & $\langle C_L \rangle$       & $std(Q_{jet}) (\times 10^4)$ \\ \midrule
\multicolumn{3}{c}{-- No Control --}        & 1.160 & 3.206 &  0.022 &  0 \\ \midrule
DRL       & 5      & 0\%   & 0.992 & 3.085 & -0.207 &  5.744 \\
DRL       & 11     & 0\%   & 0.807 & 2.958 &  0.138 & 6.526 \\
DRL       & 11     & 1\%   & 0.808 & 2.961 & -0.095 &  6.3848 \\
DRL       & 11     & 5\%   & 0.811 & 2.961 &  0.001 & 8.828 \\
DRL       & 11     & 10\%  & 0.843 & 2.982 &  0.043 & 10.382 \\ \midrule
LGPC     & 5      & 0\%   & 0.966 & 3.065 & -0.271 & 10.225 \\
LGPC     & 11     & 0\%   & 0.846 & 2.954 &  0.058 & 20.285 \\
LGPC     & 11     & 1\%   & 0.797 & 2.946 & -0.129 & 12.912 \\
LGPC     & 11     & 5\%   & 0.984 & 2.997 &  0.233 & 24.446 \\
LGPC     & 11     & 10\%  & 0.901 & 2.984 &  0.155 & 12.246 \\ \bottomrule
\end{tabular}
\caption{Summary of results.}
\label{tab:SummaryResults}
\end{table*}

\begin{table*}[t]
\centering
\begin{tabular}{>{\centering}p{0.08\linewidth-2\tabcolsep}
                >{\centering}p{0.08\linewidth-2\tabcolsep}
                >{\centering\arraybackslash}p{0.76\linewidth-2\tabcolsep}}
\toprule
Probes & Noise & Control Law ($Q_{jet} \times 10^2$)\\ \midrule
5 & 0\% &
$\begin{array}{c}  \sin\left(\tanh\left(\left( \tanh\left(\left(\left(\left( u_4\left(t-\frac{3T}{4}\right)+v_0\left(t-\frac{T}{2}\right)\right)+\left(\left(v_0+\left(u_2-u_4\left(t-\frac{T}{2}\right)\right)\right) \cdot v_3\left(t-\frac{3T}{4}\right)\right)\right)+v[1]\right)\right) \right.\right. \right. \cdot \\ 
\left.\left.\left.  \left(\left(\cos\left(\left(\left(u_4\left(t-\frac{3T}{4}\right)+v_0\left(t-\frac{T}{2}\right)\right)+\left(\left(v_0-\left(0-\left(u_2-u_4\left(t-\frac{T}{2}\right)\right)\right)\right) \right. \right.\right.\right.\right.\right.\right.\right.  \cdot \\
\left.\left.\left.\left.\left.\left.\left.\left. v_3\left(t-\frac{3T}{4}\right)\right)\right)\right)-v_2\left(t-\frac{3T}{4}\right)\right)+\left(v[2]+\left(u_4\left(t-\frac{T}{2}\right)-u_4\left(t-\frac{3T}{4}\right)\right)\right)\right)\right)\right)\right) \end{array}$  \\ \midrule
11 & 0\% & $v_6+\left(v_2-v_2\left(t-\frac{3T}{4}\right)\right)$ \\ \midrule
11 & 1\% &
$\tanh\left(\sin\left(\sin\left(v_6\right)\right)\right)-\cos\left(u_7\left(t-\frac{T}{4}\right)\right)$ \\ \midrule
11 & 5\% & 
$\tanh\left(\tanh\left(v_6\right)\right)+v_2$ \\ \midrule
11 & 10\% & $0.72965 \cdot \left( \sin\left( u_4\left(t-\frac{T}{4} \right)\right) - v_4 \left(t-\frac{T}{4}\right) \right) \cdot \left( \tanh\left(v_7\right) + \sin\left(v_6\right) \right)$ \\ \bottomrule
\end{tabular}
\caption{Control laws from LGPC.}
\label{tab:ControlLawsMLC}
\end{table*}

\section{Performance analysis}\label{s:result}

In this section, the performances of reinforcement learning and linear genetic programming are analyzed. In each training episode, the initial condition is randomly selected, thus replicating an experimental scenario, in which full control of the initial condition to start the actuation is difficult to achieve. The particular case where the starting trigger can be set corresponding to a specific case is included in Appendix A.

\subsection{Controller in the absence of noise}
\label{ss:variableIC}

\begin{figure*}
    \centering
    \includegraphics[width=0.9\linewidth]{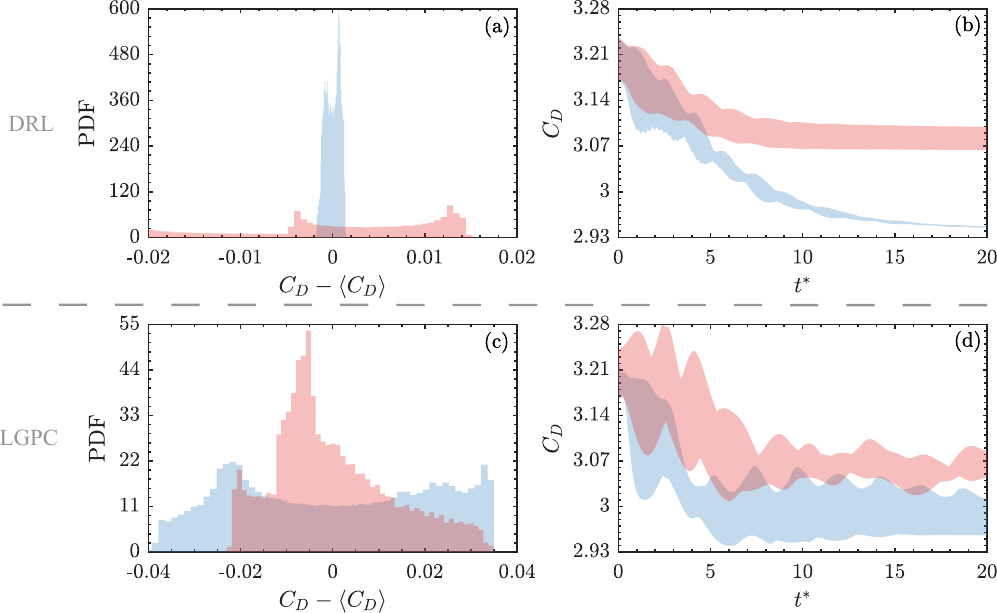}
    \caption{Influence of the initial condition on the control performance. Results are shown for DRL (a-b) and LGPC (c-d). The probability density function (a,c) and the $C_D$ envelope (b,d) are computed from the $C_D$ profiles extracted for 55 equispaced initial phases over the whole unperturbed shedding cycle. Distributions are shown for 5 \sy{redRr}{rec} and 11 \sy{blueRr}{rec} probes.}
    \label{fig:Clean_rIC}
\end{figure*}

DRL and LGPC are first trained on clean data, i.e. in absence of noise. It is important to remark that the performance of each selected actuation depends on the corresponding flow condition when the actuation is started, i.e. the same control law (or weight distribution for the ANN) determine different performances if run under different initial condition.

\begin{figure*}
    \centering
    \includegraphics[width=0.85\linewidth]{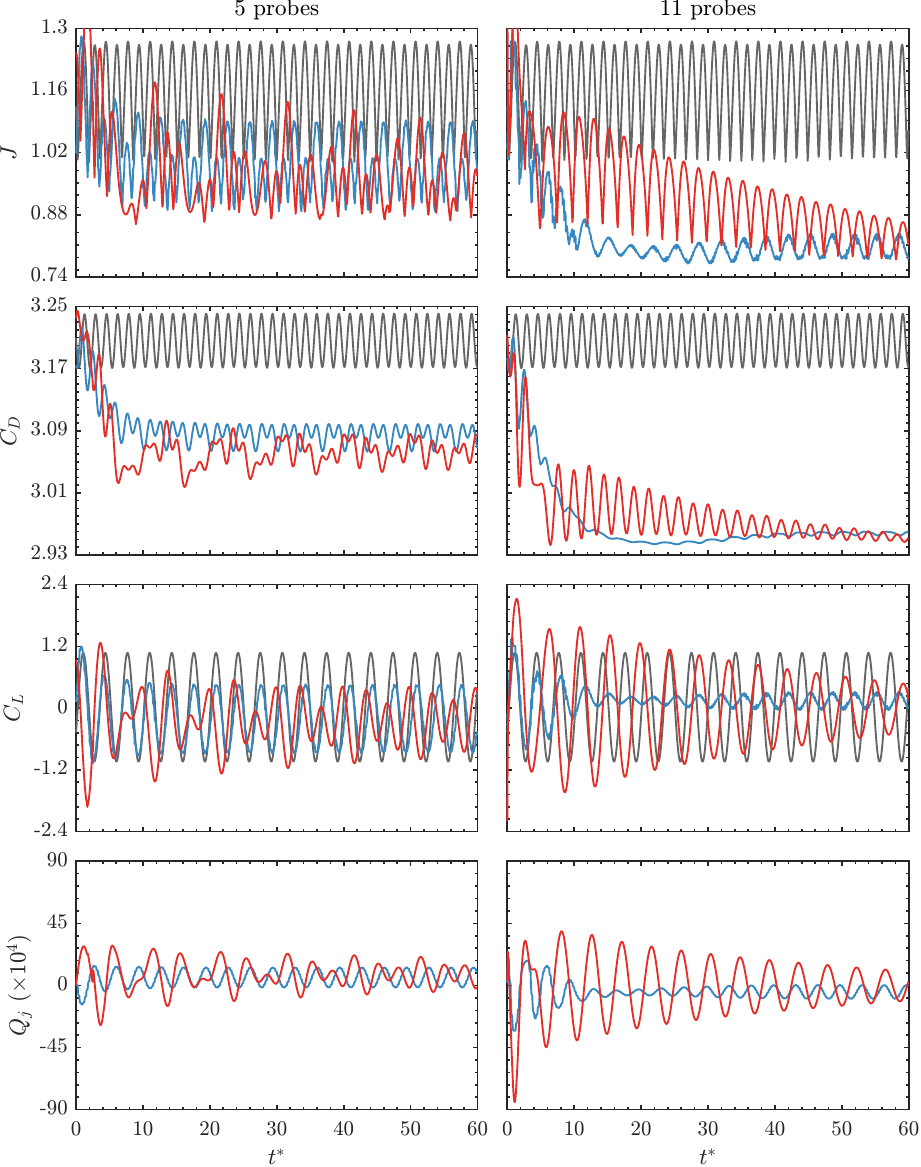}
    \caption{Evolution of $J$, $C_D$, $C_L$, and $Q_j$ upon the actuation of DRL \lcap{-}{blueR}, and LGPC \lcap{-}{redR} for 5 and 11 probes. Unforced case \lcap{-}{greyR} shown for reference.}
    \label{fig:Clean_rIC_history}
\end{figure*}

The probability distribution function (PDF) of the drag coefficient for the final selected actuation after training is illustrated in Figure \ref{fig:Clean_rIC}(a) for the case of the DRL, and in Figure \ref{fig:Clean_rIC}(c)c for LGPC, both considering 5 and 11 probes. These distributions are obtained analyzing the $C_D$ value obtained running the simulation with 55 equispaced initial phases over the whole unperturbed shedding cycle and averaging the last $650$ simulations steps, corresponding to $3.25t^*$ (being $t^* = t U_\infty/D$ and $U_\infty$ the freestream velocity), i.e. a shedding period of the unforced configuration. This is the same time interval adopted for the computation of the cost function $J$.

The PDF demonstrates that the DRL is less sensitive to the effect of the initial condition if compared to LGPC. This result is not surprising, considering that the agent of the DRL is more complex than the control laws obtained through LGPC (see Table \ref{tab:ControlLawsMLC}), thus it is potentially more flexible to variable initial conditions. While this is a desirable aspect of DRL, on the downside it comes at the expense of a less interpretable control policy. A bit more surprisingly, DRL shows a degradation in performance when passing from $11$ to $151$ probes, which is the case evaluated by \citet{rabault2019JFM}. As an hypothesis, this might be due to the non-sufficiently complex architecture of the ANN since it is common despite the sensing configuration. The input probe number is an order of magnitude larger for 151 probes, thus possibly requiring a more powerful network for the agent and more intense training. Nonetheless, it can be concluded that increasing the number of probes to a reasonable extent seems to reduce the variability of the drag coefficient, thus delivering reliable and robust action against variable initial conditions. This is also observed in Figure \ref{fig:Clean_rIC}b, where the envelope of the drag coefficient history in the set of analyzed initial conditions is presented. The shaded region is centred on the mean drag coefficient at each instant, and the half-width is set to one standard deviation of the drag coefficient. The mean values corresponding to such initial conditions are reported in Table \ref{tab:SummaryResults}.

Observing the results for LGPC, it can be observed that it benefits mainly in terms of average drag coefficient when increasing the probe number (see Figure \ref{fig:Clean_rIC}d), although there is no significant impact on the dispersion of the drag coefficient.

The time history of the cost function $J$, the drag and lift coefficients $C_D,C_L$ and the flow rate of the actuator $Q_j$, are reported in Figure \ref{fig:Clean_rIC_history} for the final selected actuation after training of DRL and LGPC. The initial condition is selected among the $50$ tested cases like the one resulting in a final drag coefficient closest to the mean value. The results are presented for $5$ and $11$ probes, including the case without actuation as a reference. 

In the limit of low probe number ($5$ probes), LGPC is observed to have slightly superior performance than DRL in terms of $C_D$. The average lift coefficient in the final phases of the observation horizon is in both cases weakly negative (i.e. LGPC and DRL converge to an asymmetric flow configuration determined by the control action). This effect is slightly more significant for LGPC, thus showing that the actuation is also aiming to alter the flow symmetry to reduce drag. In terms of the actuation flow rate, DRL converges to a substantially lower standard deviation of the flow rate of a single jet (used here as a parameter, since the net mass flow rate is zero), thus meaning that it requires less power consumption for the actuation. 

The differences are more significant for the case with $11$ probes. The actuation obtained with DRL features a faster convergence to the asymptotic drag and lift coefficient, with minimal fluctuations of the latter. This is achieved with strong action in the initial phase, which rapidly damps to a significantly lower intensity to counteract the triggering of the shedding. The actuation identified by LGCP, on the other hand, needs a significantly longer time to converge. The performances in terms of final drag coefficient are similar to DRL, although the oscillations of the lift coefficient and the flow rate of the actuators are more significant. It is nonetheless remarkable the simplicity of the obtained control law (see Table \ref{tab:ControlLawsMLC}). For the case of 11 probes with a random initial condition, LGPC converges to a law that involves only two probes, both using the crosswise velocity component, located in the middle and on one side of the cylinder (see Figure \ref{fig:NumericalSetup} for probe numbering). Remarkably, LGPC is able to identify the flow symmetry and address time-delayed feedback (for probe 2 the control law also includes a time-delayed signal with a delay of $3/4$ of the period). While the identified control law seems less robust to the effect of the initial conditions, it leads to identifying a subset of probes that is sufficient to obtain effective control.

\subsection{Robustness to measurement noise}
\label{ss:noise}

In this section, the robustness of DRL and LGPC in presence of noise is addressed. Additive noise with Gaussian distribution is included in the sensor data. Three noise levels are investigated, i.e. $1\%,5\%$ and $10\%$ of the freestream velocity.

\begin{figure*}
    \centering
    \includegraphics[width=0.9\linewidth]{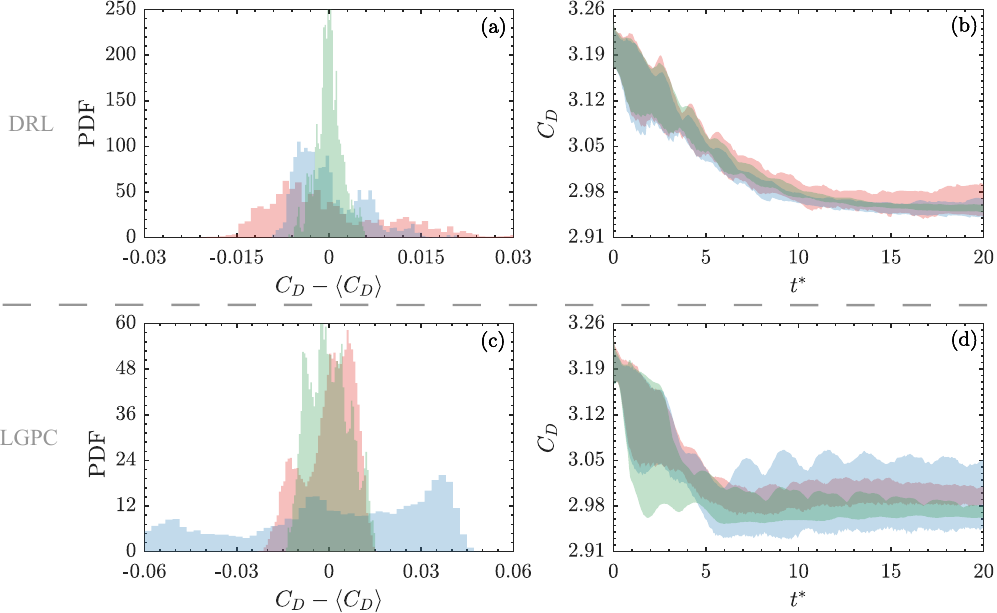}
    \caption{Influence of the initial condition on the control performance in the presence of noise. Results are shown for DRL (a-b) and LGPC (c-d). The probability density function (a,c) and the $C_D$ envelope (b,d) are computed from the $C_D$ profiles extracted for 55 equispaced initial phases over the whole unperturbed shedding cycle. Distributions are shown for 1\% \sy{greenRr}{rec}, 5\% \sy{blueRr}{rec}, and 10\% \sy{redRr}{rec} noise level.}
    \label{fig:Noise_rIC_pdf}
\end{figure*}

\begin{figure*}
    \centering
    \includegraphics[width=0.9\linewidth]{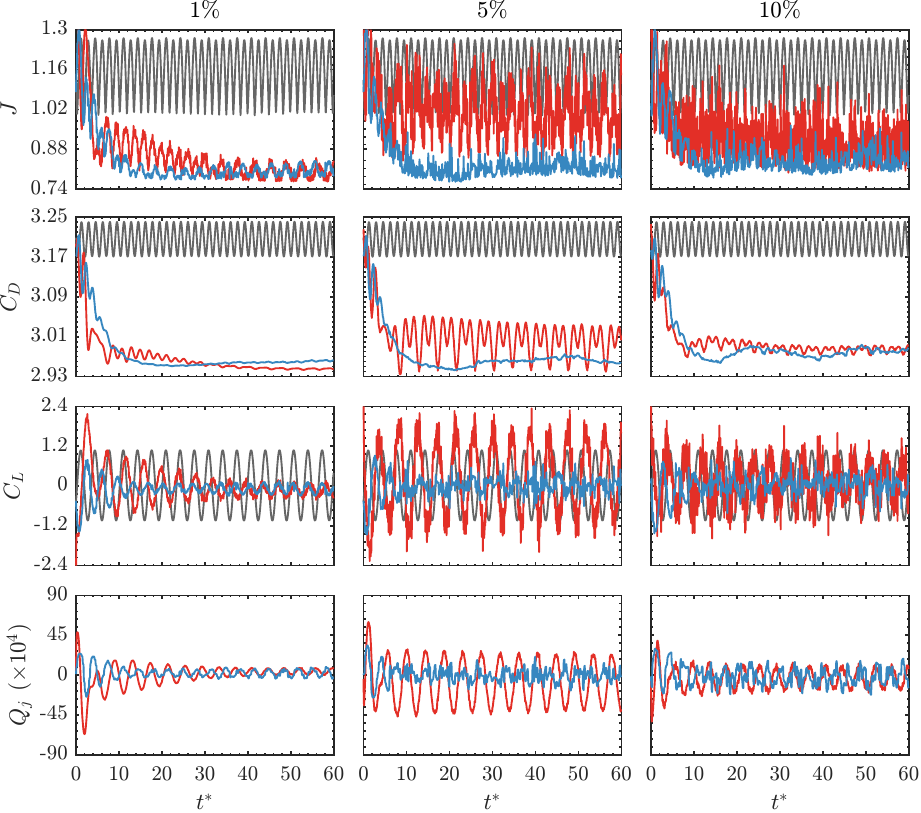}
    \caption{Evolution of $J$, $C_D$, $C_L$, and $Q_j$ upon the actuation of the DRL \lcap{-}{blueR}, and LGPC \lcap{-}{redR} controller in the presence of  1\%, 5\% and 10\% noise levels. Unforced case \lcap{-}{greyR} shown for reference.}
    \label{fig:Noise_rIC_history}
\end{figure*}

Similarly to the \S \ref{ss:variableIC}, the final actuation policy obtained after training is tested under a range of initial conditions for the case of $11$ probes. The PDF of the drag coefficient, as well as its time evolution and the corresponding dispersion for different initial conditions, are illustrated in Figure \ref{fig:Noise_rIC_pdf}a,b. For the DRL, the scatter around the mean drag coefficient is not significantly affected in the initial steps of the actuation ($t^*<5$), in which the actuation is aiming to displace the flow configuration from one limit cycle to another. For larger times, the dispersion around the mean drag coefficient seems to increase with the noise level, as expected. The final achieved performance shows only a minor degradation for noise levels up to $5\%$, while the penalty becomes more significant at $10\%$ noise level. 
Figure \ref{fig:Noise_rIC_history} reports the evolution with time of the cost function, drag and lift coefficients and actuator flow rate for different noise levels. The initial condition is chosen according to the PDF of the drag coefficient. It is set to be the one yielding the  drag coefficient most similar to the mean. It can be observed that the agent is capable to obtain in all tested cases a drag coefficient with relatively small fluctuations, while the lift coefficient experiences larger variations with the increasing noise level. This is expected since the lift coefficient is directly affected by the asymmetry introduced by the actuation, whose flow rate is directly related by the agent to the signal recorded by the probes. The drag coefficient, on the other hand, is related to the wake configuration and thus the effect is partially damped. Interestingly enough, the results reported in Table \ref{tab:SummaryResults} also show that the average lift coefficient is close to zero, i.e. the noise has the effect of avoiding the agent tricking the policy to achieve drag reduction by introducing asymmetries.

For the case of LGPC, the effect of noise appears more significant, as illustrated in Figure \ref{fig:Noise_rIC_pdf}c,d. According to the results in Table \ref{tab:SummaryResults}, the drag coefficient is quite significantly affected by increasing noise. Interestingly, for the noise level of $1\%$ and $5\%$, the obtained control law is relatively simple and still identifies that 2 probes are sufficient to perform an effective control action. In particular, probe 6 and an off-axis probe (either 2 or 7) are selected in both cases. The history of the cost function, force coefficients and actuator flow rate are also presented in Figure \ref{fig:Noise_rIC_history}. It is indeed confirmed that, for the case of low noise, the optimization successfully reduce the drag coefficient and the oscillations of the lift coefficient, with even more satisfying results than for the case without noise. For larger noise levels, the control law identified by LGPC is not capable of reducing the oscillations of the lift coefficients, thus inevitably affecting also the share of drag coefficient ascribed to vortex shedding.

A direct undesirable consequence of the simplicity of the control law is that, with the increasing noise level, the action is less effective. The different effects of noise between DRL and LGPC can be addressed on one side by the different complexity of the policy, and on the other side by the procedure to determine the action. As described in \S \ref{s:Methodology}, the action selected by the DRL agent is obtained by weighting the current ANN output with the previous step control action, thus introducing a certain soothing effect. It can be speculated that low-pass filtering of the probe signal could improve LGPC performance. Nonetheless, for fairness of comparison, we maintained the same implementation of LGPC presented originally by \citet{li2017GP}, and leave this as an object for future study.

The robustness study against noise for the 5 probes configuration leads to similar conclusions as for the 11 probes case. Hence, for the sake of brevity, it is not included. With a smaller number of probes, a higher noise is observed, as filtering out noise becomes more difficult. The probes in the wake of the cylinder have shown to be the most relevant for feedback because the vortex shedding is more pronounced and the signal-to-noise ratio is better. Intriguingly, LGPC has shown to have lower performance degradation than DRL when using only 5 probes. This is not surprising, since all the control laws extracted with LGPC in the 11 probe case lead to parsimonious use of probes, rarely exceeding 3 or 4 sensors.

\begin{figure*}[t]
    \centering
    \includegraphics[width=0.9\linewidth]{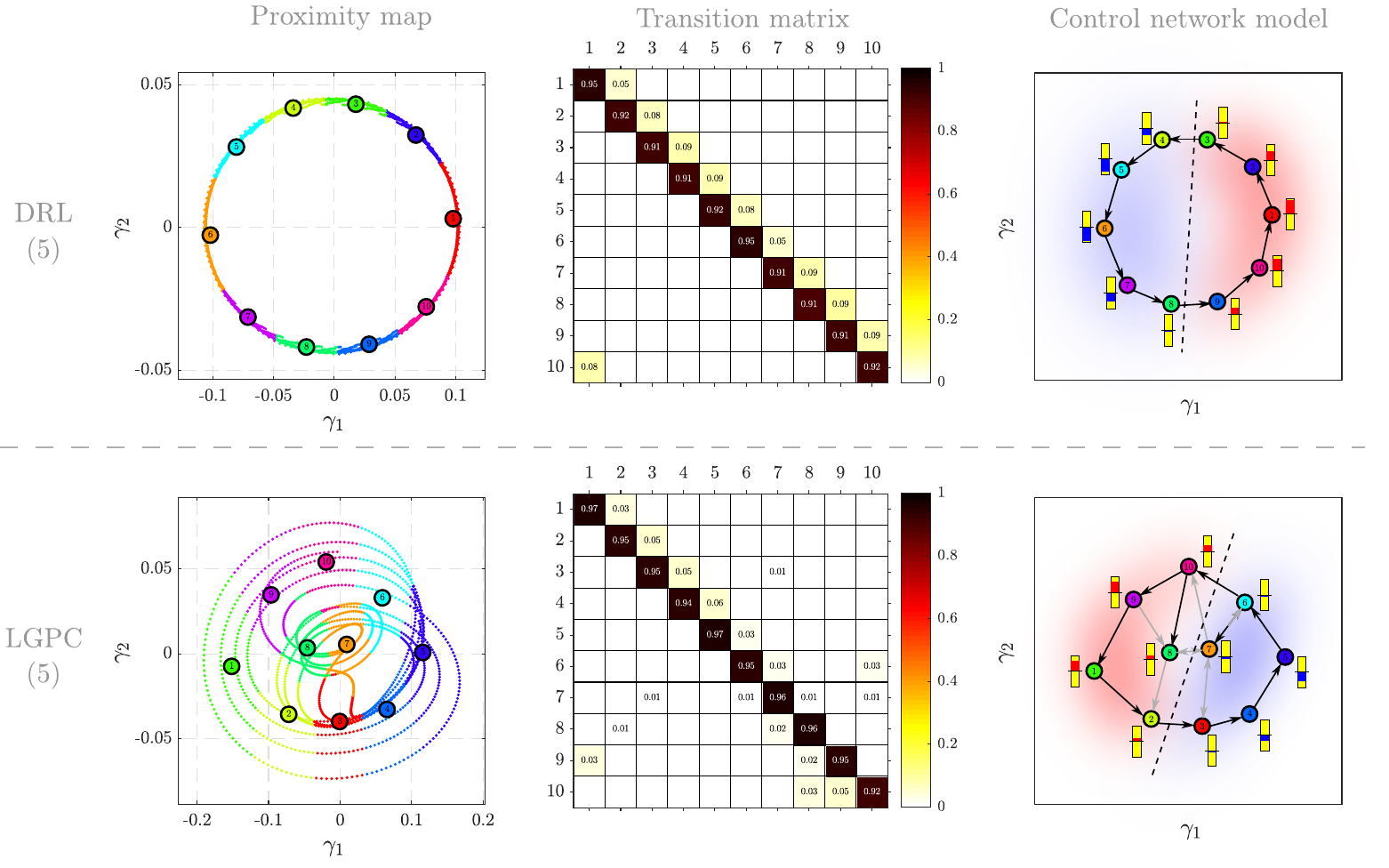}
    \caption{Visualization of the laws/policies learned by DRL (top) and LGPC (bottom) for the 5-sensor configuration. (left) The proximity maps of the sensor signals of the controlled regime show each cluster in a different colour with their centroids denoted by numbers (1-10). (centre) The transition matrix is associated with the clustering process. (right) The control network. The yellow boxes set the maximum range of $b$ during the control and the red and blue boxes indicates positive and negative levels (the sign of $b$). The dashed line and red/blue background indicate the assumed separation between the actuation regions.} \label{fig:CL_S5}
\end{figure*}
\begin{figure}[h!]
    \centering
    \includegraphics[width=0.99\linewidth]{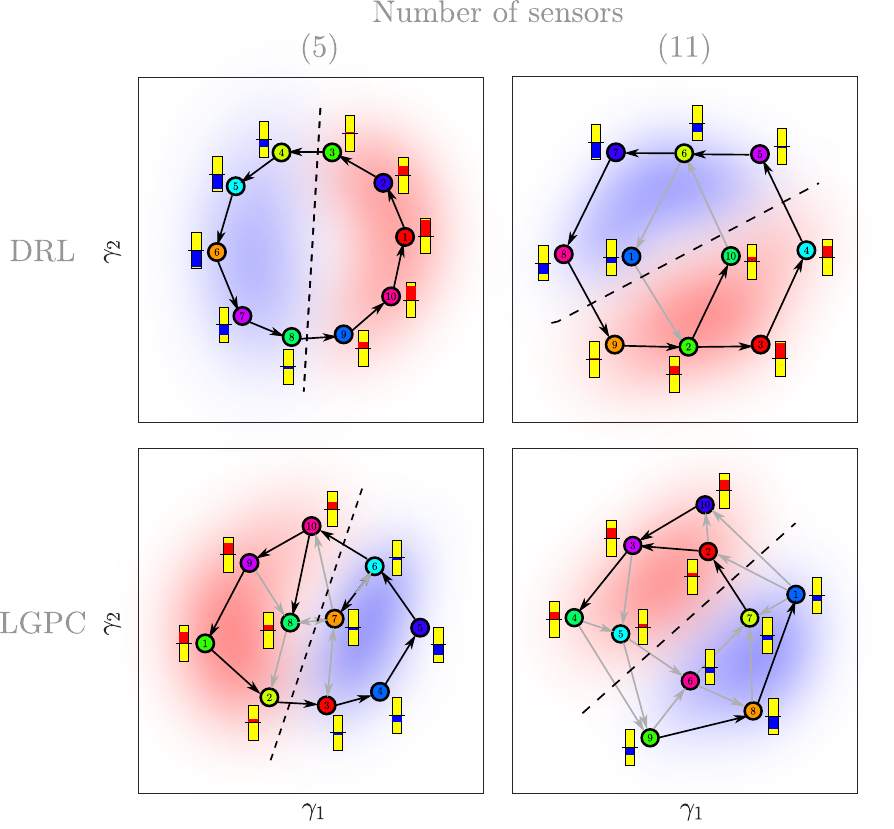}
    \caption{Control network for laws/policies learned by DRL (top) and LGPC (bottom) for the 5 (left) and 11 (right) probes. For more details see the caption of Figure~\ref{fig:CL_S5}}
    \label{fig:CL_S}
\end{figure}

\section{Control results and discussion \label{s:interpretability}}
In this section, the controls achieved by DRL and LGPC are analysed with the clustering method described in \S~\ref{Sec:Method_ControlLawVisu}.

\begin{figure*}[t]
    \centering
    \includegraphics[width=0.9\linewidth]{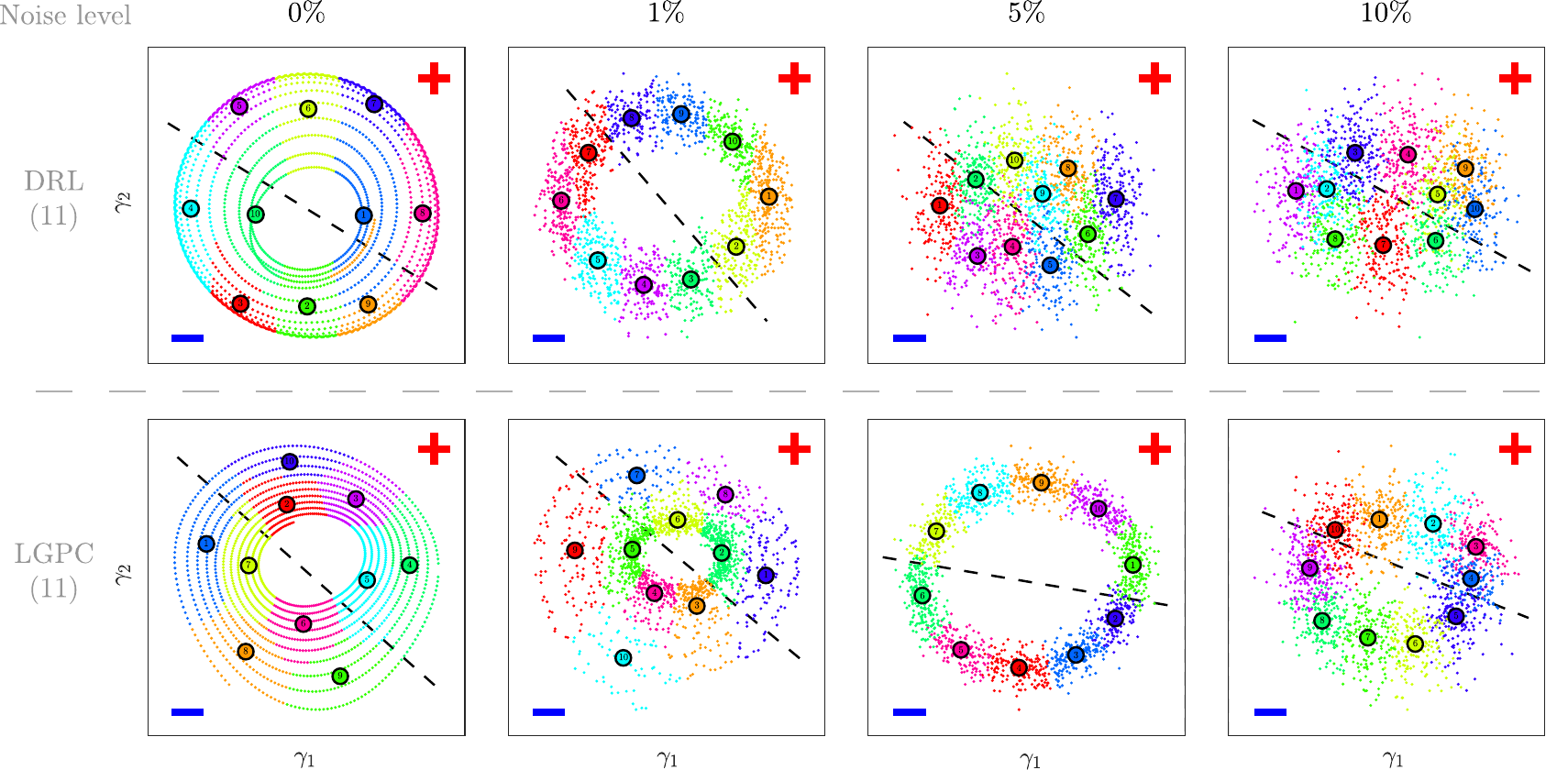}
    \caption{Proximity map of the clustered sensors and centroids for the velocity time series extracted from the DRL/LGPC controls with increasing noise. The noise intensity increases from left to right: $0\%$, $1\%$, $5\%$ and $10\%$. The dashed black line divides the reconstructed phase space into two regions: one of positive actuation and one with negative actuation (the sign of $b$) denoted respectively by a red $+$ and a blue $-$. The $\gamma_1$ and $\gamma_2$ axis have been reflected such the red $+$ is located in the top right corner for all cases.}
    \label{fig:CL_Noise}
\end{figure*}

For all cases the transient has been excluded by the analysis by removing the first $3000$ time steps, corresponding to $15t^*$. It must be remarked, however, that for the LGPC cases with 0\% and 1\%-level noise, the transient lasts for more than the observed $60$ time units.

Figure~\ref{fig:CL_S5} shows the cluster-based interpretation methodology applied to the laws/policies learned by DRL and LGPC for the 5-sensor cases without noise. The interpretation methodology is applied independently to each case. First, note that the proximity maps of the sensor time series are reminiscent of phase space. Such visualization is not surprising as the controlled flow remains mainly periodic and therefore can be represented in a 2D space. The transition matrices display high probabilities for the self-transitions (diagonal values). This is due to the high sampling rate that excessively populates each centroid. The transition states are then under-represented, hence the low values of the transition probabilities outside the diagonal. Thus, the sensor time series have been subsampled to 1/5th of the data to artificially increase the transition probabilities located outside the diagonal. This operation is done to render the transition matrix easier to read while it does not change either the results or the interpretations. For the DRL control case, note that for each cluster there is only one arrival cluster. This is translated in the control network by a cycle that can be interpreted as a limit cycle in the phase space. For the LGPC case, a limit cycle can also be inferred from the data. However, two centroids close to the centre of the limit cycle (centroids 7 and 8) play a role in short-circuiting the limit cycle. Concerning the control achieved, in both cases, the plane is divided into two regions: one of positive actuation including half the centroids and one of negative actuation including the other half. Furthermore, note that the actuation level increases with the distance to the dividing line. Such organization of the actuation around a limit cycle indicates that the actuation mechanism employed for drag reduction is phasor control, meaning that the control depends on the flow phase.

In addition to this visualization, an analysis combining the two dynamics (not included here for brevity) is performed. For this, the time series have been appended and 20 centroids have been chosen for the description of the flow. The resulting proximity map shows the two limit cycles on the same plane, almost concentric, revealing that the controlled flows are close dynamically. However, the probability transition matrix loses the dynamic relationship between the centroids as some clusters contain states from both learning strategies. 

Figure~\ref{fig:CL_S}, displays the control network for both the number of sensors (5 and 11). Note that the overall shape and relations between the centroids remain the same when going from $5$ sensors to $11$ sensors. Again, the main limit cycle is divided into two regions of positive and negative actuation revealing, again, a phasor control mechanism.

Finally, the impact of noise on the actuation mechanism is investigated. Figure~\ref{fig:CL_Noise} displays time series and the centroids projected in the proximity map for the controlled flows with increasing noise. First, note that going from $0\%$ to $1\%$, the noise disturbs the smooth distribution of data. For DRL, as the noise level increases the clusters starts to overlap so much that for the $10\%$-level noise case, separating visually the clusters becomes impossible on the proximity map. The resulting control network, not plotted because of its complexity, shows that for the post-transient regime the dynamics are mainly driven by the noise. Nonetheless, note that the centroids are distributed around an ellipsoid, indicating a periodicity in the flow. As for the LGPC cases, the proximity maps describe successfully the controlled dynamics. For 0\% and 1\%-noise level, the centroids describe a spiral which is consistent with the long transients (see corresponding lift coefficient $C_L$ in figures~\ref{fig:Noise_rIC_history} and \ref{fig:Noise_rIC_pdf}). For the 5\%-noise level, the persistent high-amplitude oscillations plotted in figure~\ref{fig:Noise_rIC_history} are represented by a clear limit cycle in the proximity map and for the 10\%-noise level, the centroids describe a cycle blurred by the noise which is consistent with the low-amplitude oscillations of the lift coefficient in the post-transient regime (figure~\ref{fig:Noise_rIC_history}). In summary, in all the cases, the flow includes a periodic behaviour. Regarding the actuation command, all the proximity maps can be separated into two regions, one of positive actuation and negative actuation, indicating that both DRL and LGPC managed to learn a phasor control regardless of the noise level.

\section{Conclusions}\label{s:Conclusions}

Two machine-learning-based strategies which minimize the drag of a cylinder exhibiting vortex shedding wake are evaluated. The performance of Deep Reinforcement Learning and Linear Genetic Programming Control in identifying effective control policies has been assessed in the realistic conditions of a limited number of sensors and noise contamination of the sensed data. The training is performed using random initial conditions, in the attempt to reproduce an experimental scenario in which it is not feasible to control the full flow state when the actuation is started.

It is observed that, in absence of noise, the average performance achieved by DRL and LGPC is similar in terms of average lift and drag coefficient, although with a significant advantage of the DRL in terms of the standard deviation of the mass flow rate of the actuation. It must be remarked, however, that the amount of power needed to control has not been included in the loss function. Furthermore, the policy obtained by DRL appears to be more robust in terms of dependency on the initial condition. On the other hand, LGPC achieves significantly more compact and interpretable control laws, which also identify only subsets of the probes as being relevant to define control actions. In particular, for the cases with 11 probes and low to moderate noise levels, LGPC identifies that only 2 probes are sufficient to define a sufficiently robust control law. The potential reduction in sensorization complexity is a very desirable feature for experimental application. 

Regarding the effect of noise, DRL shows superior performances in terms of robustness to noise of the sensors up to relatively high noise levels (10\%). LGPC, on the other hand, is able to identify control laws that are effective in reducing the average drag coefficient, although maintaining a larger level of fluctuations around the mean lift and drag coefficient if compared to DRL, and in general larger standard deviation of the actuator mass flow rate. This superior robustness of the DRL can certainly be ascribed to the higher complexity of the adopted agent if compared to the control policies identified by LGPC. Also in presence of noise, LGPC converges to compact control laws and automatically identifies only a few significant probes for the control, almost independently of the noise level within the tested range.

Finally, an analysis using clustering of the sensor data using MDS has been carried out to interpret the control laws obtained by DRL and LGPC. In absence of noise, it is rather evident the convergence to a limit cycle in both cases and a clear relation between the phase of the shedding and the control actuation. This is an indicator that both solutions converge to phasor control, which was an expected result for this simple flow configuration. 

Although the number of sensors might seem high for a real application, it is remarkable that this study has already considerably reduced the number of probes compared to other recent contributions~\citep{Ren2021pof,rabault2019DRL,rabault2019JFM}. Having a smaller set of probes is viable but not recommended for a proof of concept as the one presented herein, since the information provided to the controller would be very limited and hence the solution would probably be suboptimal. The chosen sets of sensors allow testing the performance of both algorithms and also evaluate their relevance for the final controller. Consequently, one of the main conclusions of the study is the capability of LGPC of identifying a subset of probes as the most relevant, which suggests the possibility of further reduction. This is a milestone for future implementations in a real-world application or experiment.

Any comparison contains subjective biases associated with the computational load, the number of parameters, the complexity of the control problem, and even the experience of authors with various approaches. Also, each approach could have been improved. e.g., DRL has many architectures with different performances and LGPC consistently profits from subplex iterations\cite{cornejo2021gMLC}. Even, the very formulation of the control ansatz will affect the performance. Yet, our study points already to desirable features of two different machine learning approaches. Future machine learning control can be expected to integrate the fast adaptive learning of DRL, the analytical laws and interpretability of LGPC, the fast optimization of cluster-based control for smooth control laws\cite{Nair2019jfm} and the mathematically rigorous framework of Bayesian optimization\cite{Blanchard2022ams}, to name only a few aspects.

\begin{acknowledgments}
Work produced with the support of a 2020 Leonardo Grant for Researchers and Cultural Creators, BBVA Foundation, grant n. IN[20]\_ING\_ING\_0163. The Foundation takes no responsibility for the opinions, statements and contents of this project, which are entirely the responsibility of its authors.

Funding of the National Natural Science Foundation China (NSFC) under grants 12172109 and 12172111 and a Natural Science \& Engineering grant of the Guangdong province, China, is gratefully acknowledged. \end{acknowledgments}

\section*{Data Availability Statement}
The data that support the findings of this study are available from the corresponding author upon reasonable request.
\appendix
\section{Flow control performance training at fixed Initial Condition}
\label{ss:fixedIC}

\begin{figure}
    \centering
    \includegraphics[width=0.9\linewidth]{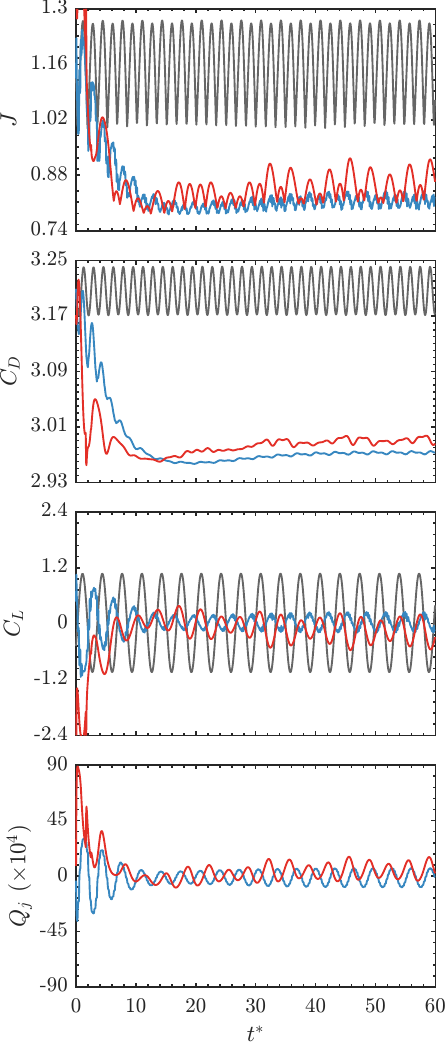}
    \caption{Evolution of $J$, $C_D$, $C_L$, and $Q_j$ for the unforced case \lcap{-}{greyR}, RL \lcap{-}{blueR}, and LGPC \lcap{-}{redR} at fixed initial condition during the training process. Results are shown 11 probes.}
    \label{fig:Clean_fIC}
\end{figure}

The case with training starting from a fixed initial condition is analyzed here. Although this condition is difficult (if not impossible) to achieve in experiments, it can be approximated reasonably well in the case of shedding-dominated flows at a low-to-moderate Reynolds number. The evolution of the cost function, the force coefficients and the actuator flow rate are reported in Figure \ref{fig:Clean_fIC} for the case of the final resulting actuation after the training. For brevity, only the case with 11 probes is analyzed.

\begin{figure*}
    \centering
    \includegraphics[width=0.85\linewidth]{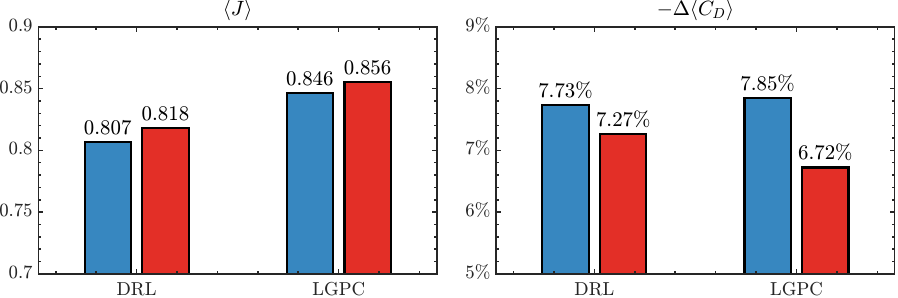}
    \caption{Key performance indicators, cost function $\langle J \rangle$ (left) and drag reduction $-\Delta \langle C_D \rangle = \frac{\langle C_{D_0} \rangle - \langle C_{D_0} \rangle}{\langle C_{D} \rangle}$ (right), for random \sy{blueR}{rec} and fixed \sy{redR}{rec} initial condition.}
    \label{fig:barplot}
\end{figure*}

Since in this case the initial condition for DRL and LGPC is the same, it is interesting to compare the different strategies adopted by the two control laws to bring the system to a different limit cycle than the original unperturbed system. In the first shedding cycle, the control law identified by LGPC acts with a strong blowing on the top actuator, thus determining a significant asymmetric and a net negative lift coefficient. On the other hand, DRL acts with a quasi-periodic change in phase in opposition to the original shedding cycle.

While the performance in terms of drag coefficient is similar for DRL and LGPC, more significant differences arise in the lift coefficient. The control law identified by DRL reduces more significantly the oscillation of the $C_L$, thus also requiring a smaller actuator flow rate for the control action. The bar plot of the cost function and of the drag coefficient reduction in figure~\ref{fig:barplot} further support this assertion. For both LPGC and DRL, a reduction in performances when passing from random to fixed initial condition is observed. This is a first glance surprising; it is hypothesized that having a random initial condition aids the exploration in both algorithms, allowing the discovery of more robust control strategies.

Surprisingly, even if using the same set of probes, LGPC exhibits a much more complex control law if compared to the random initial condition case, and fails to identify a compact law,

\begin{equation*}
    \begin{split}
        Q_{jet} = 10^{-2} \times \left[
        \left(\left(u_7\left(t-\frac{T}{2}\right) \cdot \left(v_6+u_6\left(t-\frac{3T}{4}\right)\right)\right)+ \right.\right. \\
        \left.\left. \left(0.1551-\left(u_8\left(t-\frac{3T}{4}\right) \cdot \left(v_0\left(t-\frac{T}{4}\right) \right.\right.\right.\right.\right.+ \\
        \left.\left.\left.\left.\left. \left(v_9\left(t-\frac{T}{2}\right)-u_0\left(t-\frac{T}{4}\right)\right)\right)\right)\right)\right)*cos\left(u_7\left(t-\frac{T}{2}\right)\right)\right]
    \end{split}
\end{equation*}\\

This might be ascribed to a tendency of the algorithm to overfit the control law to the prescribed initial condition.
\section*{References}
\bibliography{bibliography.bib}

\end{document}